\documentclass[preprint, 10pt]{sigplanconf}

\usepackage{url}
\usepackage{epsfig}
\usepackage{amsmath}

\begin{document}

\conferenceinfo{WXYZ '05}{date, City.}
\copyrightyear{2005}
\copyrightdata{[to be supplied]}

\titlebanner{banner above paper title}        
\preprintfooter{short description of paper}   

\title{Parallelizing Stream with Future}

\authorinfo{Rapha\"el Jolly}
           {Databeans, V\'elizy-Villacoublay, France}
           {raphael.jolly@free.fr}

\maketitle

\begin{abstract}
Stream is re-interpreted in terms of a Lazy monad. 
Future is substituted for Lazy in the obtained construct, 
resulting in possible parallelization of any algorithm 
expressible as a Stream computation. The principle is 
tested against two example algorithms. Performance is 
evaluated, and a way to improve it briefly discussed.
\end{abstract}



\section{Introduction}

\begin{figure}
\centering
\epsfig{file=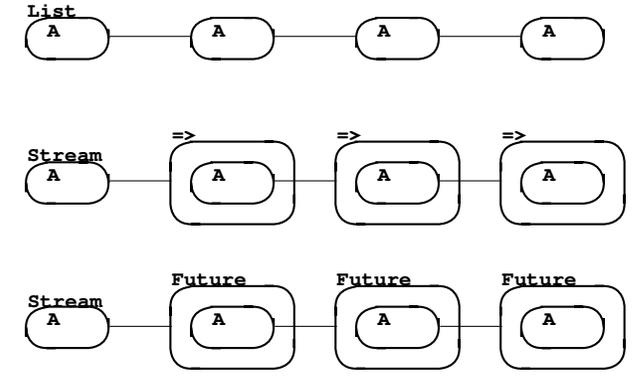,clip=,width=0.95\linewidth}
\caption{Chaining of elementary cells}
\label{fig:chaining}
\end{figure}

Scala parallel collections \cite{Prokopec:2011} exploit data or SIMD
parallelism, whereby a unique operation is applied in parallel to several
data, independantly. There are problems however, where sub-parts are not
independant. In such cases, some sequence must be re-introduced, to allow
certain tasks to operate only after some others have ended. This is called
task parallelism or pipe-line. To achieve it, if we do not want to descend
to thread level, one alternate option is to use a message passing scheme,
such as the one implemented in Scala in the form of Future \cite{Haller:2011}.
We seek to assemble futures in a way that allows us to obtain parallelization
of some suitable algorithms. Let us take the Stream concept of a lazily
evaluated List as a model. List is implemented as a chain of elementary cells:

\begin{verbatim}
class Cons(hd: A, tl: List[A])
    extends List[A]
\end{verbatim}

In Stream, tail is evaluated lazily, using a by-name parameter:

\begin{verbatim}
class Cons(hd: A, tl: => Stream[A])
    extends Stream[A]
\end{verbatim}

If, instead of waiting for the moment when it is requested, \texttt{tail}
starts to compute itself asynchronously on a new thread, we obtain a parallel
computation. The elementary cell must be modified as:

\begin{verbatim}
class Cons(hd: A, tl: Future[Stream[A]])
    extends Stream[A]
\end{verbatim}

This is illustrated in Figure \ref{fig:chaining}. This idea should allow
us to parallelize any algorithm that can be expressed functionally and
recursively as a Stream.

\section{Outline}

The paper is organized as follows : in Section \ref{sec:monad}, we will
introduce a Lazy monad that is semantically equivalent to the pass-by-name
parameter used in Stream elementary cells. In Section \ref{sec:stream},
we will explain how to rewrite Stream in terms of this monad, or any other
one for that matter, namely the Future monad. In Section \ref{sec:prime} and
\ref{sec:mult} we will test our parallelization scheme against two example
algorithms. Lastly, in Section \ref{sec:eval} we will discuss our results
and suggest some directions of improvement.

\section{Lazy monad}
\label{sec:monad}

We examine a construct that behaves like \texttt{=> A} and at the same
time obeys the monad rules. As an illustration, we take the example of
the \texttt{Traversable.filter} method. In List, it is implemented in
an imperative, iterative way:

\begin{verbatim}
def filter(p: A => Boolean): List[A] = {
  val b = new ListBuffer[A]
  for (x <- this) if (p(x)) b += x
  b.result
}
\end{verbatim}

Functionally, this would have to be expressed recursively:

\begin{verbatim}
def filter(p: A => Boolean): List[A] = {
  var rest = this
  while (!rest.isEmpty && !p(rest.head))
    rest = rest.tail
  rest match {
    case head::tail => head::tail.filter(p)
    case Nil => Nil
  }
}
\end{verbatim}

But it requires as many stack frames as elements in the List,
resulting in stack overflows (tail call optimization is not applicable
here because \texttt{filter} is not the last operation in the method body).
In Stream, this is avoided by the pass-by-name nature of the second parameter
to \texttt{cons}, allowing \texttt{filter} not to be called again immediately,
and the number of stack frames to stay below reasonnable level. As a result,
the computation is not performed immediately but on an on-demand basis.

\begin{verbatim}
def filter(p: A => Boolean): Stream[A] = {
...
  if (!rest.isEmpty)
    cons(rest.head, rest.tail filter p)
  else Empty
}
\end{verbatim}

To achieve the same behavior with a monad, we use again an extractor
as in the List case, but we suppose that its second member is not
forced, i.e. it is of type \texttt{=> A}. Then we suppose that we can
transform this type \texttt{=> A} through a (for now putative) method
\texttt{map}.

\begin{verbatim}
def filter(p: A => Boolean): Stream[A] = {
...
  rest match {
    case head#::tail =>
      head#::tail.map(_ filter p))
    case Empty => Empty
  }
}
\end{verbatim}

Likewise, we require the second parameter of the constructor \texttt{\#::}
to be by-name, as the laziness is to be forwarded by \texttt{map}. Let us now
sketch the form of our Lazy monad. In order to ease later substitutions with
Future, let us name it the same.

\begin{verbatim}
trait Future[+A] extends (() => A) {
  def map[B](f: A => B) = Future(f(apply()))
  def flatMap[B](f: A => Future[B]) =
      f(apply())
}
object Future {
  def apply[A](value: => A) = new Future[A] {
    lazy val apply = value
  }
}
\end{verbatim}

We find that our construct has type \texttt{() => A} and a method \texttt{map},
both as expected. We lastly endow a method to force its value, in a similar
fashion as Future, for the reason given above.

\begin{verbatim}
object Await {
  def result[A](future: Future[A],
      duration: Duration) = future()
}
\end{verbatim}

\section{Stream re-interpretation}
\label{sec:stream}

Every method of Stream can be rewritten in the same spirit. Let us skip
the details, and concentrate on the implementation of elementary Cons cells.
In List, the constructor's second parameter is a normal, ``flat'' type. The
extractor provided by the case class gives us back this value as-is.

\begin{verbatim}
case class ::[A](hd: A, tl:
    List[A]) extends List[A] {
  override def isEmpty: Boolean = false
  override def head: A = hd
  override def tail: List[A] = tl
}
\end{verbatim}

Conversely, in Stream it is a by-name parameter. Since case classes disallow
such a parameter type, the Cons cell must be a normal class, and no extractor
is provided. Calls to \texttt{tail} force the value, which is memoized.

\begin{verbatim}
object Stream {
  class Cons[+A](hd: A, tl:
      => Stream[A]) extends Stream[A] {
    private[this] var tlVal: Stream[A] = _
...
    def tailDefined: Boolean = tlVal ne null
    override def tail: Stream[A] = {
      if (!tailDefined) tlVal = tl
      tlVal
    }
  }
}
\end{verbatim}

Our monad-based implementation is as follows. The second parameter of the
constructor is a monad. Calls to \texttt{tail} force the value as above.
Extractions however do not, and give us back the genuine monad, thus
preserving the laziness. When forced, memoization of the value occurs
internally and needs not be done again in the Cons cell.

\begin{verbatim}
object Stream {
  case class Cons[+A](hd: A, tl:
      Future[Stream[A]]) extends Stream[A] {
    private[this] var defined: Boolean = _
...
    def tailDefined = defined
    override def tail: Stream[A] = {
      defined = true
      Await.result(tl, Duration.Inf)
    }
  }
}
\end{verbatim}

Below we give the methods to get our modified Stream back and forth from/to
the original Scala Stream. These are implemented recursively like the other
modified Stream methods. Notice the call to \texttt{future} made in
\texttt{apply} in order to wrap tails into their monadic containers. The
reverse operation in \texttt{unapply} is simply done by forcing the value
through calling \texttt{tail}.

\begin{verbatim}
object Stream {
  def apply[A](s: scala.Stream[A]):
      Stream[A] =
    if (s.isEmpty) Empty
    else cons(s.head, future(apply(s.tail)))
  def unapply[A](s: Stream[A]):
      Option[scala.Stream[A]] = Some(
    if (s.isEmpty) scala.Stream.Empty
    else scala.Stream.cons(s.head,
        unapply(s.tail).get))
}
\end{verbatim}

\section{Example : prime sieve}
\label{sec:prime}

To evaluate our parallel algorithm, we have first tested it against a prime
sieve \cite{Scala:2003}. It is not the most efficient, as it scans every
divisors of a number up to the number itself instead of just its square
root, but it turns out to be parallelizable accoring to our technique.
First we give the original algorithm with the normal Stream implementation.

\begin{verbatim}
val n = 20000
def primes = sieve(Stream.from(2))
def sieve(s: Stream[Int]): Stream[Int] =
    Stream.cons(s.head,
    sieve(s.tail.filter { _ % s.head != 0 }))
primes.take(n).force
\end{verbatim}

The modified implementation of Stream entails the following modifications
to the example code : use an extractor to obtain head and (wrapped) tail ;
call map on tail to express further operations.

\begin{verbatim}
def primes = sieve(Stream.range(2, n, 1))
def sieve(s: Stream[Int]): Stream[Int] =
    s match {
  case head#::tail =>
    head#::tail.map(s =>
    sieve(s.filter { _ % head != 0 }))
  case Empty => Empty
}
primes.force
\end{verbatim}

The purpose of \texttt{force} is to wait for the computation to complete.
Notice that we have defined the desired number of terms in advance, otherwise
the computation will not stop since it is asynchronous (if Future is used ;
in the case of Lazy, the behavior is the same as with normal Stream).

\section{Example : polynomial muliplication}
\label{sec:mult}

The second example that we have tested our scheme against, is a computer
algebraic algorithm of sparse polynomial multiplication. Other researches
and applications of streaming algorithms for such kind of computations
can be found in \cite{Neun:1989, Schwab:1992, Kredel:1994}. We use
multivariate polynomials, in distributive representation:

\[x={x}_{0}+{x}_{1}+\mathrm{...}+{x}_{n}\]
\[{x}_{\mathrm{i}}={c}_{\mathrm{i}}{m}_{\mathrm{i}}\]

The test case, detailed in \cite{Fateman:2002}, simply consists in computing
the product of two such big polynomials:

\[xy\]

Decomposing polynomial multiplication into a sequence of
multiply-by-a-term-and-add operations, it is possible to express
the algorithm in terms of a stream computation.

\begin{verbatim}
type T = Stream[(Array[N], C)]
def times(x: T, y: T) = (zero /: y) {
    (l, r) =>
  val (a, b) = r
  l + multiply(x, a, b)
}
\end{verbatim}

Multiply-by-a-term is expressed functionnaly/recursively as follows.

\begin{verbatim}
def multiply(x: T, m: Array[N], c: C) =
    x match {
  case (s, a)#::tail => {
    val (sm, ac) = (s * m, a * c)
    val result =
      (sm, ac)#::tail.map(multiply(_, m, c))
    if (!ac.isZero) result
    else result.tail
  }
  case Empty => Empty
}
\end{verbatim}

\begin{figure}
\centering
\epsfig{file=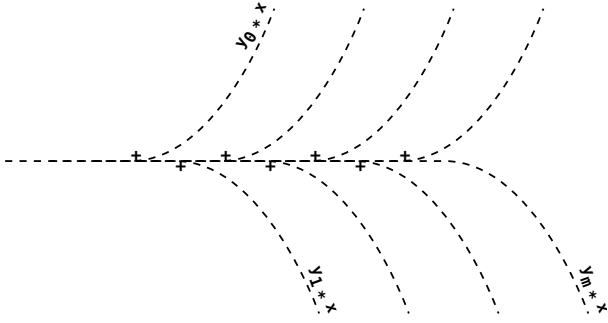,clip=,width=0.95\linewidth}
\caption{Streaming multiply and add operations}
\label{fig:streaming}
\end{figure}

Polynomial addition is also implemented recursively. Note that the tail has
to be forced in the case when one term cancels, which results in a call to
Await.result. This is not considered good in a regular use of Futures, but
we have not been able to avoid it (and it does not occur all the time).
Figure \ref{fig:streaming} illustrates the process.

\begin{verbatim}
def plus(x: T, y: T) = x match {
  case (s, a)#::tailx => y match {
    case (t, b)#::taily => {
      if (s > t)
        (s, a)#::tailx.map(plus(_, y))
      else if (s < t)
        (t, b)#::taily.map(plus(x, _))
      else {
        val c = a + b
        val result =
            (s, c)#::(for (sx <- tailx;
                           sy <- taily)
            yield plus(sx, sy))
        if (!c.isZero) result
        else result.tail
      }
    }
    case Empty => x
  }
  case Empty => y
}
\end{verbatim}

\section{Evaluation}
\label{sec:eval}

\begin{figure}
\centering
\epsfig{file=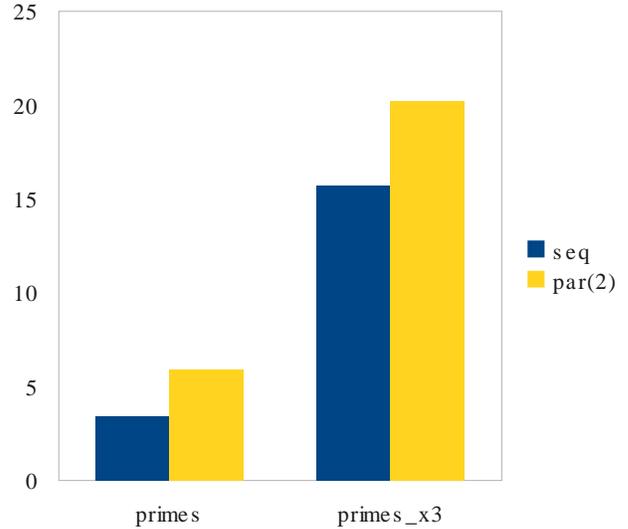,clip=,width=0.95\linewidth}
\caption{Timings for primes (seconds)}
\label{fig:chart}
\end{figure}

\begin{figure}
\centering
\epsfig{file=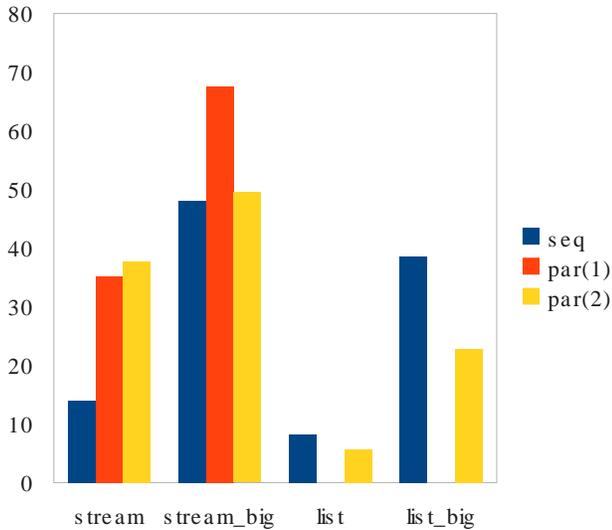,clip=,width=0.95\linewidth}
\caption{Timings for polynomial multiplication (seconds)}
\label{fig:timings}
\end{figure}

To evaluate our method, we have run the examples both in sequential and
parallel mode (using Lazy and Future respectively). Computations were performed
on a single core Intel Atom D410 with hyperthreading and 2GB memory, under Linux
version 2.6.32-5-amd64 (Debian 6.0) with java version ``1.7.0\_17'', OpenJDK
64-Bit Server VM (mixed mode) and scala-2.11.0-M2. The primes example was run
in two versions, \texttt{primes} and \texttt{primes\_x3}, until number 20000 and
60000 respectively. The polynomial multiplication example was also run in two
versions, \texttt{stream} and \texttt{stream\_big}, the latter using polynomials
with bigger coefficients (of a factor 10000000001), in order to increase the
footprint of elementary operations. According for instance to \cite{Chen:2002},
the expected speedup with hyperthreading should be on the order on 1.20. This
is what we obtain with a control computation, \texttt{list} (and
\texttt{list\_big}), which uses a more classical parallelization technique,
based on parallel collections \cite{Jolly:2013}. Our results are presented
in Table \ref{timings} and Figure \ref{fig:chart} and \ref{fig:timings}.
On the vertical axis, \texttt{seq} means sequential execution, \texttt{par(1)}
means parallel execution with available processors set to 1, and \texttt{par(2)}
means normal parallel execution on our platform.

\begin{table}[!t]
\renewcommand{\arraystretch}{1.3}
\caption{Timings (seconds)}
\label{timings}
\centering
\begin{tabular}{|c||c|c|c|}
\hline
            &  seq & par(1) & par(2) \\
\hline\hline
primes      &  3.4 &        &    5.9 \\
\hline
primes\_x3  & 15.7 &        &   20.2 \\
\hline\hline
stream      & 14   &   35.1 &   37.7 \\
\hline
stream\_big & 48   &   67.5 &   49.5 \\
\hline
list        &  8.2 &        &    5.7 \\
\hline
list\_big   & 38.6 &        &   22.7 \\
\hline
\end{tabular}
\end{table}

We make the following observations:

\begin{enumerate}
\item scaling does not occur in the primes example, probably due to too
fine-grained elementary operations
\item the polynomial multiplication example does not scale either in the
small coefficient version
\item the streaming approach, at least in the polynomial example, seems to
be sound, and perform reasonably well when no parallelization is involved
(\texttt{stream} is not worse than half as fast than \texttt{list}, which
is a well optimized classical iterative/imperative implementation)
\item the overhead incurred by parallelization, well visible when available
processors is set to 1, is compensated when the footprint of coefficients
is big enough, as in \texttt{stream\_big}, and performance increases
consistently with what we can expect of hyperthreading
\end{enumerate}

As a way to improve our technique, since the minimum size of elementary
computations seems to be a key factor, we suppose that grouping these in
bigger chunks may provide better efficiency. This will have to be tested
in forthcoming research.

\section{Conclusion}

We have presented a technique for parallelizing algorithms expressible as
stream computations. Stream was rewritten in terms of a Lazy monad, which
was then replaced by Future, enabling parallel execution of computation
subparts. Two applications were proposed, for prime numbers computation
and polynomial multiplication, respectively. Evaluation showed that
this method has an overhead, but that it can scale nonetheless if
elementary computations are big enough, even on a limited platform
such as a hyperthreaded mono-processor.


\bibliographystyle{abbrvnat}
\bibliography{lazy}


\end{document}